\documentclass[journal]{IEEEtran}
\usepackage{graphicx,amsmath,subfigure,epstopdf,color}
\usepackage{amsmath,graphicx,siunitx}

\pdfoutput=1 
\usepackage{ifpdf}
\begin{document}
%
\title{Estimation of Self-Field Critical Current and Transport-Magnetization AC Losses of Roebel Cables}
%
%
%

\author{Francesco~Grilli, Michal~Vojen\v ciak, Anna~Kario, and Victor~Zerme\~no
\thanks{F. Grilli, A. Kario and V.  Zerme\~no are with the Karlsruhe Institute of Technology, Karlsruhe, Germany. M.  Vojen\v ciak is with the Slovak Academy of Science, Bratislava, Slovakia. Corresponding author's e-mail: francesco.grilli@kit.edu.}
\thanks{Manuscript received October 16, 2015.}}

\maketitle

\begin{abstract}
Roebel cables made of HTS coated conductors are regarded as promising cables for winding applications in virtue of their large engineering current density and low losses. The composing meander-shaped strands are assembled very tightly into the cable, which results in a strong electromagnetic interaction between them. This interaction profoundly influences the effective self-field critical current ($I_c$) of the cable, which is much lower than the sum of the $I_c$s of the composing strands. Also the AC losses are influenced by the material's properties and by the geometrical configuration of the cable. Being able to predict the effective critical current and AC losses of such cables is very important for a proper design of applications: due to the complexity of the cable's geometry and of the material's properties, this prediction can only be performed with advanced numerical tools.

In this contribution we use finite-element-based models to compute the effective $I_c$ and the AC losses of Roebel cables composed of 31 strands using tapes from two manufacturers. The AC losses are analyzed in the simultaneous presence of transport current and background perpendicular field proportional to the current, which mirrors the situation occurring in a winding. Our models include the angular dependence of $J_c(B,\theta)$ at~\SI{77}{\kelvin}, which is very different for the two materials. 

By means of a successful comparison of the simulation results to experimental data obtained with a calorimetric method measuring the evaporation of liquid nitrogen, this work confirms the applicability and efficiency of our numerical techniques for simulating the electromagnetic behavior of Roebel cables and devices thereof.
\end{abstract}
\section{Introduction}
\IEEEPARstart{T}{he} 
idea of assembling coated conductors in a Roebel cable shape was proposed ten years ago~\cite{Goldacker:CEC-ICMC05}. Now these cables are manufactured in a variety of configurations and offer great potential for applications such as magnets, transformers, and electrical machines, where cables with large engineering current density, good mechanical properties and low AC losses are needed~\cite{Goldacker:SST14}.

Until now the AC losses of Roebel cables have been measured mostly in the case of transport current only ~\cite{Jiang:SST10, Terzieva:SST10, Lakshmi:TAS11, Jiang:SST12} or applied magnetic field only~\cite{Terzieva:SST10, Lakshmi:SST10b, Lakshmi:TAS11, Majoros:TAS14}. Only in~\cite{Jiang:SST13} were the losses caused by the simultaneous application of an AC transport  current and AC background magnetic field measured. In that work, the losses of a Roebel cable composed of six \SI{2}{\milli\meter}-wide strands with a self-field critical current of~\SI{212}{\ampere} (calculated value) at~\SI{77}{\kelvin} were measured with the standard electrical technique.

One issue of measuring the total losses with electrical technique is the possible interference of the various signals and the extraction of the `true' loss signal, which in general requires very careful shaping and positioning of the voltage taps~\cite{Vojenciak:JPCS06}. For this reason we chose here a calorimetric technique, which measures the quantity of evaporated nitrogen caused by the AC losses.

The samples under investigation are high-current Roebel cables subjected to in-phase transport current and background magnetic field (perpendicular to the flat face of the cables). The field was varied proportionally to the transport current, to match what happens in superconducting windings.

The focus of this contribution is the estimation of the self-field critical current and total AC losses of the cable by means of numerical models, which are able to take into account the geometric layout of the cable as well as the complex $J_c(B,\theta)$ dependence of the superconductor materials of the manufacturers

\section{Properties of the Roebel Cables}

\begin{figure}[t!]
\begin{center}
\includegraphics [width=\columnwidth]{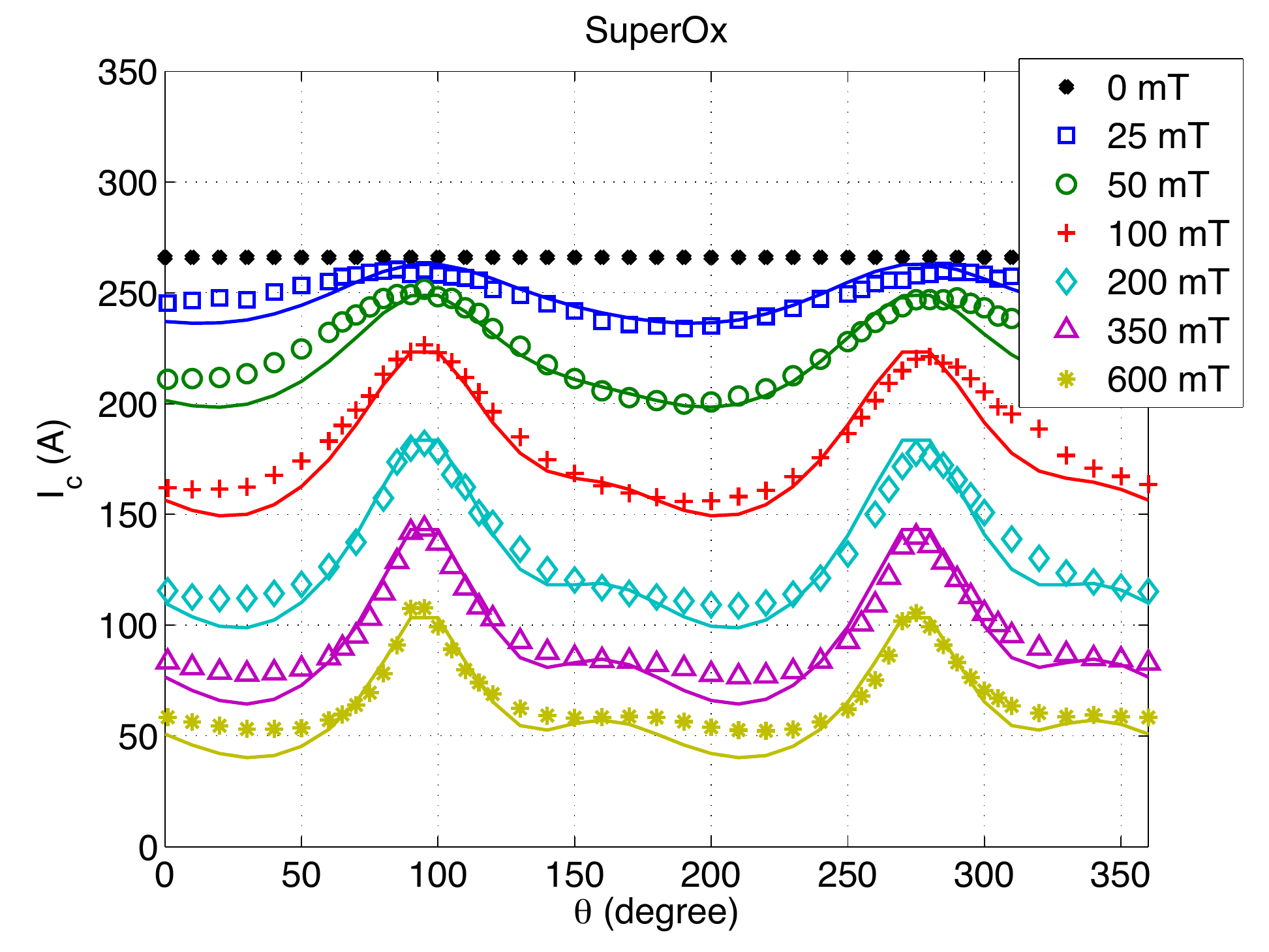}
\end{center}
\caption{\label{fig:IcH_SOx}Angular dependence of $I_c$ of the SuperOx sample. The continuous lines indicate the $I_c$ values calculated with the method described in the text. The self-field critical current is \SI{266}{\ampere}.}
\end{figure}

\begin{figure}[h!]
\begin{center}
\includegraphics [width=\columnwidth]{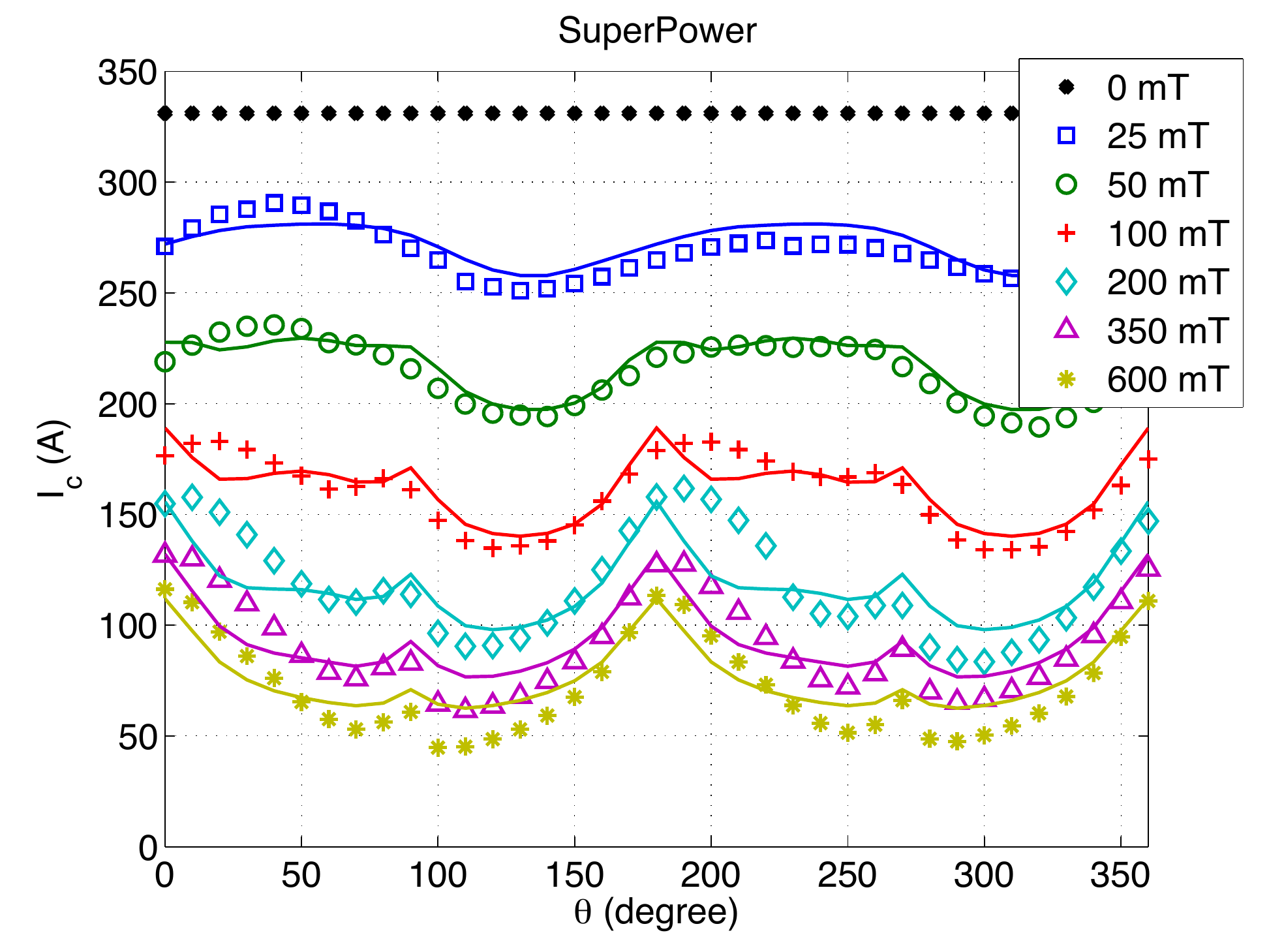}
\end{center}
\caption{\label{fig:IcH_SP}Angular dependence of $I_c$ of the SuperPower sample. The continuous lines indicate the $I_c$ values calculated with the method described in the text. The self-field critical current is \SI{331}{\ampere}.}
\end{figure}

Two Roebel cable samples were assembled from \SI{12}{\milli\meter}-wide HTS coated conductor tape from two manufactures: SuperOx~\cite{Company:SuperOx} and SuperPower, Inc.~\cite{Company:Superpower}.
The tapes have different self-field critical currents and angular dependencies, as shown in Figs.~\ref{fig:IcH_SOx} and \ref{fig:IcH_SP}. In particular, the critical current of the SuperOx tape presents two clear peaks when the field is parallel to the tape (angle equal to \SI{90}{\degree} and \SI{270}{\degree} in the figure). On the other hand, the SuperPower tape is characterized by the presence of artificial pinning centers, which flattens the angular dependence of $I_c$ and shifts the positions of the peaks. 

After punching, the SuperOx and SuperPower strands have an average critical current of \SI{129}{\ampere} and \SI{137}{\ampere}, and a $n$ power index of $33$ and $23$, respectively.

Both cables are composed of 31 \SI{5.5}{\milli\meter}-wide strands, with a measured self-field critical current at~\SI{77}{\kelvin} exceeding \SI{2}{\kilo\ampere}. The transposition length is \SI{426}{\milli\meter} and the final length of the cable about \SI{1}{m}.

\begin{figure}[t!]
\begin{center}
\includegraphics [width=\columnwidth]{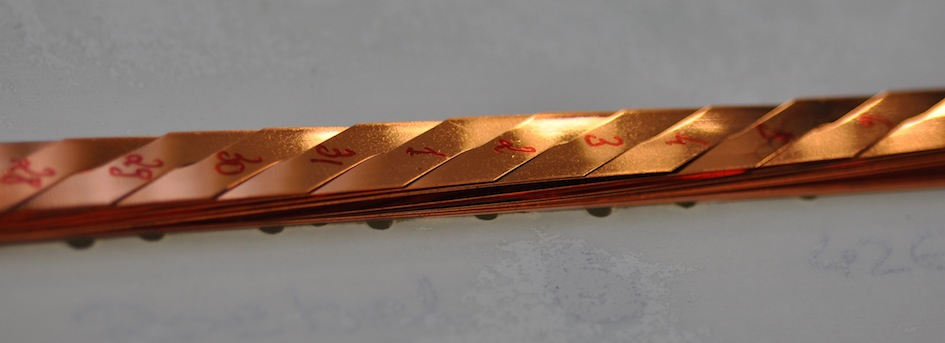}
\end{center}
\caption{\label{fig:Roebel_426_31strands}Photograph of a Roebel cable composed of 31 strands. The width of the cable is \SI{12}{mm}. The transposition length of the strands is \SI{426}{\milli\meter}.}
\end{figure}


\section{Numerical Models} 
The models for calculating the critical current and the AC losses of the cable need a $J_c(B,\theta)$ as input. This can be extracted from the measured angular dependence of $I_c$ as follows~\cite{Pardo:SST11, Grilli:TAS14c}. Based on the appearance of the data (e.g. number and position of the peaks, rate of decrease of $I_c$ with increasing field amplitude), a functional $J_c(B,\theta)$ dependence is chosen. Then the angular dependence of $I_c$ as a function of the applied field is calculated taking into account the effect of the self-field. A set of parameters that gives a reasonably good reproduction of the data is finally selected. This procedure is not a simple fit of the experimental data of Figs.~\ref{fig:IcH_SOx} and \ref{fig:IcH_SP}, but it involves the use of a model to calculate the effects of the self-field. In this case, as in \cite{Grilli:TAS14c}, we chose the finite-element magnetostatic model described in~\cite{Gomory:SST06b}.

Once $J_c(B,\theta)$ is known, we estimate the self-field critical current of the cable following two approaches. The first consists in repeating essentially the same type of magnetostatic calculation described above, but for the cable geometry instead of the individual tape and with no external magnetic field (self-field conditions)~\cite{Vojenciak:SST11}. This approach represents a situation of good current sharing, because it corresponds to assuming that each point in the superconducting strands is at the local critical current density (which varies with the field amplitude and orientation, and consequently with the position). It is based on the critical state model and labeled CSM in Tab.~\ref{tab:Ics}. 

The second approach uses a power-law $E-J$ relationship for the superconductor 
\begin{equation}\label{eq:PL}
\rho(J)=\frac{E_c}{J_c(B,\theta)} \left | \frac{J}{J_c(B,\theta)}\right | ^ {n-1}
\end{equation}
and is based on the asymptotic limit of Faraday's equation when time approaches infinity~\cite{Zermeno:SST15}. Publicly available codes (both open-source and commercial) of this method have recently been published~\cite{Zermeno:arxiv15}. This method allows for the definition of two criteria for the critical current: (i) the current at which the voltage drop per unit length has reached its critical value $E_c$ in at least one conductor, or (ii) the current at which the average voltage drop per unit length has reached its critical value $E_c$. In Tab.~\ref{tab:Ics}, these criteria are labeled MAX and AVG, respectively. 

For the calculation of AC losses, a 2-D finite-element model based on the $H$-formulation of Maxwell's equations is followed~\cite{Brambilla:SST07}. The 2-D simplification is justified by the fact that the cross-over region of the strands is much shorter than the transposition length~\cite{Terzieva:SST10, Jiang:SST10}. The cross-section of the Roebel cable is modeled as two stacks of rectangular tapes, representing the superconductor material~\cite{Grilli:SST10b, Thakur:SST11b, Pardo:SST12a}, with one additional tape on top to account for the odd number of strands.

An AC transport current $I_0 \sin(\omega t)$ is imposed by means of current constraints, a background magnetic field $B_0 \sin (w t)$ by means of Dirichlet boundary conditions~\cite{Brambilla:SST07}. The amplitudes of the magnetic field and of the transport current are linked by a proportionality constant, $B_0=\alpha I_0$. This constant $\alpha$ assumes the following values: 0, 33, 66, \SI{100}{\milli\tesla\per\ampere}. The frequency of current and field was \SI{72}{\hertz}.

\section{Experimental Set-up}
All the measurements were performed in liquid nitrogen at atmospheric pressure. The critical current was experimentally measured with the standard four-point technique and \SI{1}{\micro\volt\per\centi\meter} criterion over at least one transposition length. 

The AC losses was measured calorimetrically, by monitoring the flow of evaporated nitrogen caused by the power dissipation~\cite{Murphy:TAS13,Kario:CCA14}. The cable was placed in a racetrack coil about \SI{50}{\centi\meter} long, so that the uniformity zone of the field covered one transposition length. A bubble catcher of the same length was placed on top of the cable, and the flow of evaporated nitrogen was sent to a flowmeter. A calibration was necessary in order to link the measured nitrogen flow to the dissipated power. This was done by means of a resistive heater of known DC dissipation, placed on top of the cable. Full details of the experimental technique will be given in an upcoming publication of ours.

\section{Results}
This section presents the simulation results for critical current and AC losses, and a comparison with the experiments.
\subsection{Critical Current}\label{sec:critical_current}
Table~\ref{tab:Ics} summarizes the measured and computed critical currents of the two cable samples. The column labeled `Design' indicates the values obtained by multiplying the average self-field critical current of the strands ({\SI{129}{\ampere} and \SI{137}{\ampere} for SuperOx and SuperPower samples, respectively) by the number of strands. A comparison of the design and measured values provides information about the effects of the self-field. Interestingly, the SuperPower strands have higher self-field critical current (as also visible from the \SI{0}{\milli\tesla} curve of the original tapes in Figs.
\ref{fig:IcH_SOx} and \ref{fig:IcH_SP}), but the SuperPower cable has lower self-field critical current than its SuperOx counterpart. This is because the SuperPower tapes have a steeper degradation of their transport properties in applied field than the SuperOx ones, and the fact that they start from a higher self-field value is not sufficient to counterbalance this strong decrease. By comparing the \SI{100}{\milli\tesla} curves in Figs.
\ref{fig:IcH_SOx} and \ref{fig:IcH_SP}, one can see that the critical current of the SuperOx sample never goes below \SI{150}{\ampere} and in a certain range of angles it  exceeds \SI{200}{\ampere}. This does not happen for the SuperPower tapes, whose critical current never exceeds \SI{183}{\ampere} and goes below \SI{150}{\ampere} at certain angles.

  \begin{table}[h!]
    \begin{center}
    \renewcommand{\arraystretch}{1.8}
            \caption{Measured and Computed Critical Currents (A)}
    \label{tab:Ics}
    \begin{tabular}{l  l  l l l l}
Sample & Measured & Design & CSM & AVG & MAX \\ \hline 
SuperOx		& 2747	& 3999 & 2558	& 2509	& 2420 \\
SuperPower	& 2264 	& 4247 & 2460	& 2241	& 2090
    \end{tabular}
    \end{center}
    \label{}
\end{table}%

In fact, finite-element calculations show that the self-field of these Roebel  cables is in the range of 100-\SI{150}{\milli\tesla}, which explains why the critical current of the SuperPower cable is lower. The self-field reduction of the two cables (with respect to the design value) is \SI{31}{\percent} \SI{47}{\percent}, respectively.

\begin{figure}[t!]
\begin{center}
\includegraphics [width=\columnwidth]{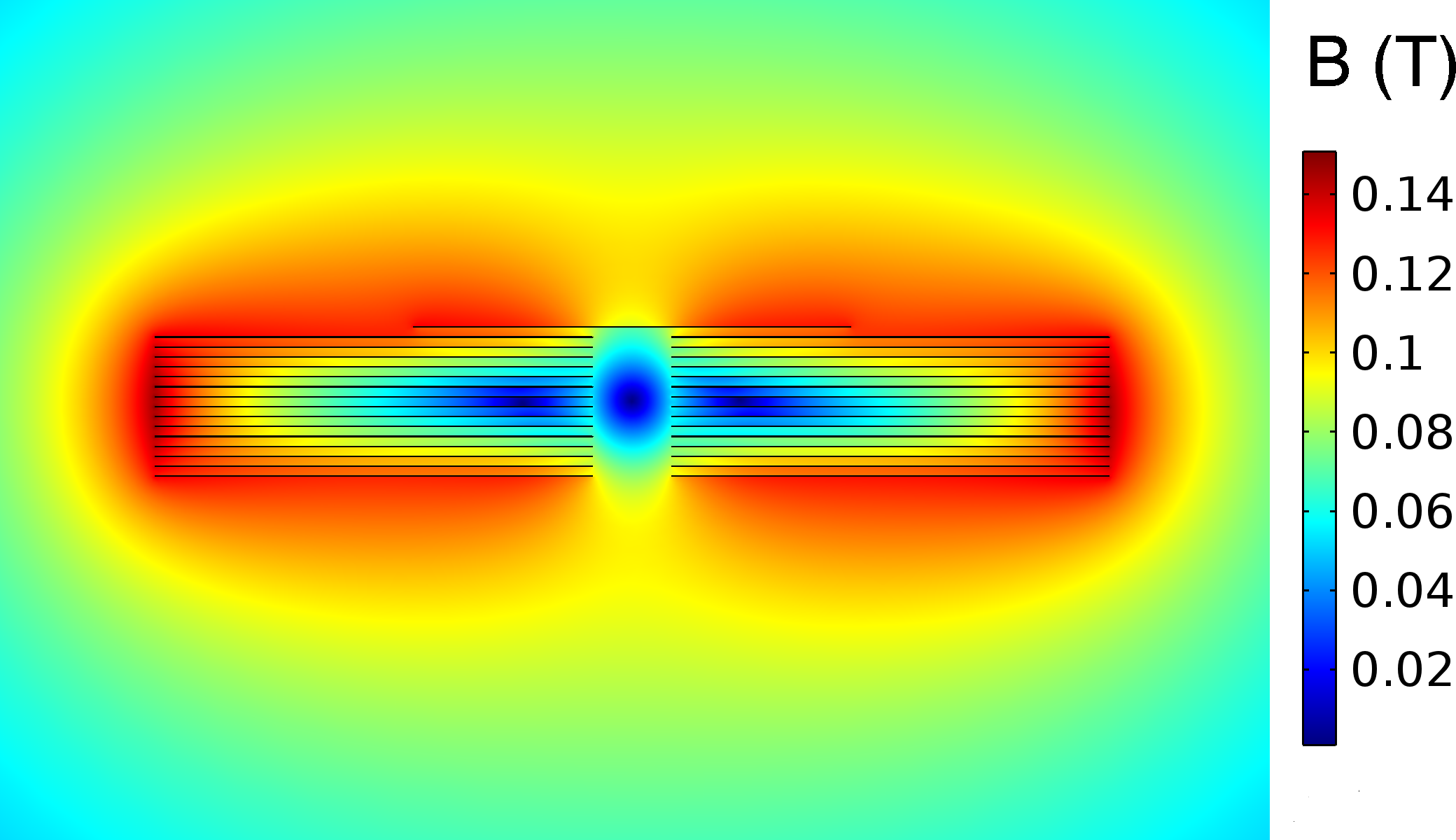}
\end{center}
\caption{\label{fig:SuperPower_Bm_MV}Self-field magnetic flux density distribution in the cross-section of the SuperPower cable carrying its critical current. The calculation is performed with the CSM approach mentioned in the text.}
\end{figure}


Overall both the CSM and power-law AVG methods predict critical currents within \SI{9}{\percent} of the experimental value. This can be considered satisfactory, given the different factors that could have contributed to this discrepancy, including statistical variations of the critical currents of the strands.

The MAX criterion makes more conservative estimates of $I_c$~\cite{Zermeno:SST15}. Its `local' definition of $I_c$ is different from the one used in experiments, where the voltage drop is averages because the voltage taps cover more than a transposition length, and it is inserted in the table just for comparison purpose. 

\subsection{AC Losses}
The time-dependent model used for computing AC losses provides the current density and magnetic field profiles at all the instants of the AC cycle.  
Two examples of those profiles are shown in Figs.~\ref{fig:Bevol} and \ref{fig:Jevol} for the SuperPower cable with a transport current of \SI{1697}{\ampere} (\SI{1200}{\ampere} r.m.s.) and a field of \SI{56}{\milli\tesla} (corresponding to $\alpha$=\SI{33}{\micro\tesla\per\ampere}). Due to the interaction of transport current and applied field, the obtained profiles are asymmetric with respect to the center of the cable.
This is in contrast with the cases of transport current only or background magnetic field only~\cite{Grilli:SST10b}.
In particular, at the peak of the AC cycle ($\omega t=3\pi/2$), the magnetic field generated by the transport current has different signs in the right and left halves of the cable's cross section, so that one one side it adds to the background magnetic field, whereas on the other side it subtracts from it. This is illustrated in Fig.~\ref{fig:Bevol}, where it is shown that the magnetic flux density in the right part of the cable is as high as \SI{120}{\milli\tesla}, whereas the left part of the cable has a maximum field of \SI{70}{\milli\tesla} in a very limited area.
This asymmetric behavior is also present in the current density profiles (Fig.~\ref{fig:Jevol}), where one can observe that current penetrates much farther in the right part of the cable than in the left one.

The interaction between the transport current and the background field strongly influences the losses. Figures~\ref{fig:TotalLosses_SuperOx} and \ref{fig:TotalLosses_SuperPower} display the total losses as a function of transport current for different values of the proportionality constant of the applied field. Results from calorimetric measurements and finite-element calculations are shown. The AC losses greatly increase in the presence of a background field (more than one order of magnitude with respect to the transport current only case). 

In general, numerical simulation successfully reproduce the experimental trend. The predicted values are within \SI{40}{\percent} of the measured ones.
These results can be considered satisfactory given all the possible sources of error, which, in addition to those mentioned at the end of section~\ref{sec:critical_current}, include also those related to the calorimetric measurement technique. The experimental data generally present a lower slope than the calculated ones, and the cause of this is not fully understood yet. Our upcoming paper on the will contain a detailed discussion of the experimental issues. 

\begin{figure}[t!]
	\centering
			\includegraphics[width=\columnwidth]{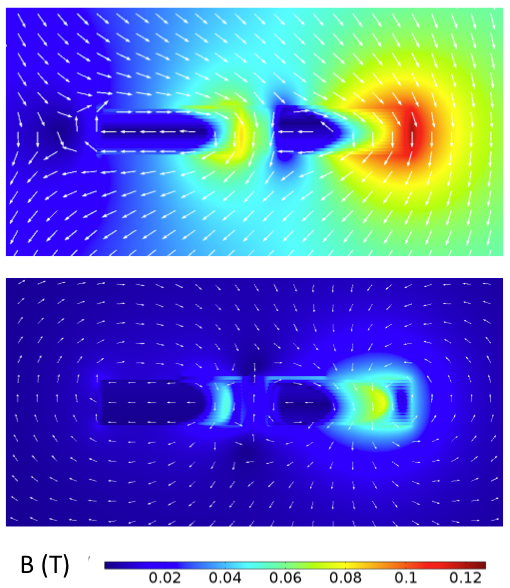}
	\caption{	\label{fig:Bevol}Magnetic flux density distribution in the SuperPower cable at selected instants of the sinusoidal AC current and field: $\omega t = 3\pi/2$ (top) and $\omega t =2\pi$ (bottom).}	
\end{figure}

\begin{figure}[t!]
	\centering
			\includegraphics[width=\columnwidth]{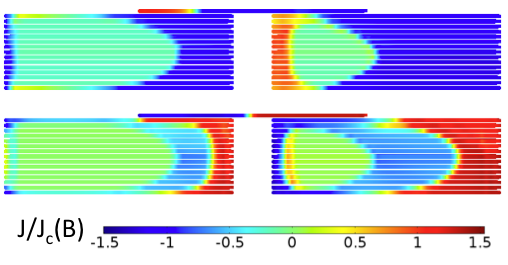}
	\caption{	\label{fig:Jevol}Current density distribution in the SuperPower cable at selected instants of the sinusoidal AC current and field: $\omega t = 3\pi/2$ (top) and $\omega t =2\pi$ (bottom).}	
\end{figure}

\begin{figure}[t!]
\begin{center}
\includegraphics [width=\columnwidth]{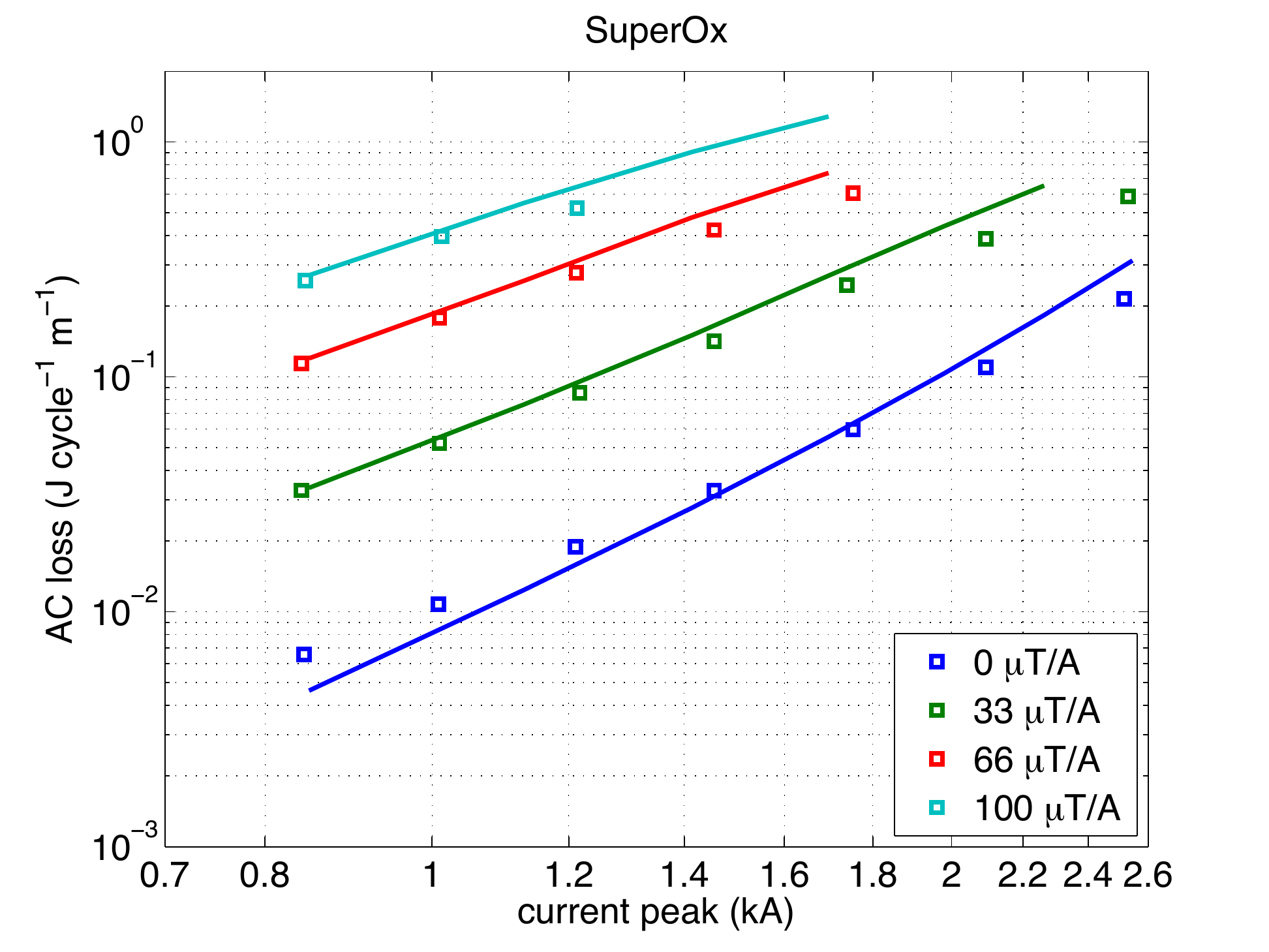}
\end{center}
\caption{\label{fig:TotalLosses_SuperOx}Measured and calculated AC losses of the SuperOx cable as a function of the transport current and for different background fields.}
\end{figure}
\begin{figure}[h!]
\begin{center}
\includegraphics [width=\columnwidth]{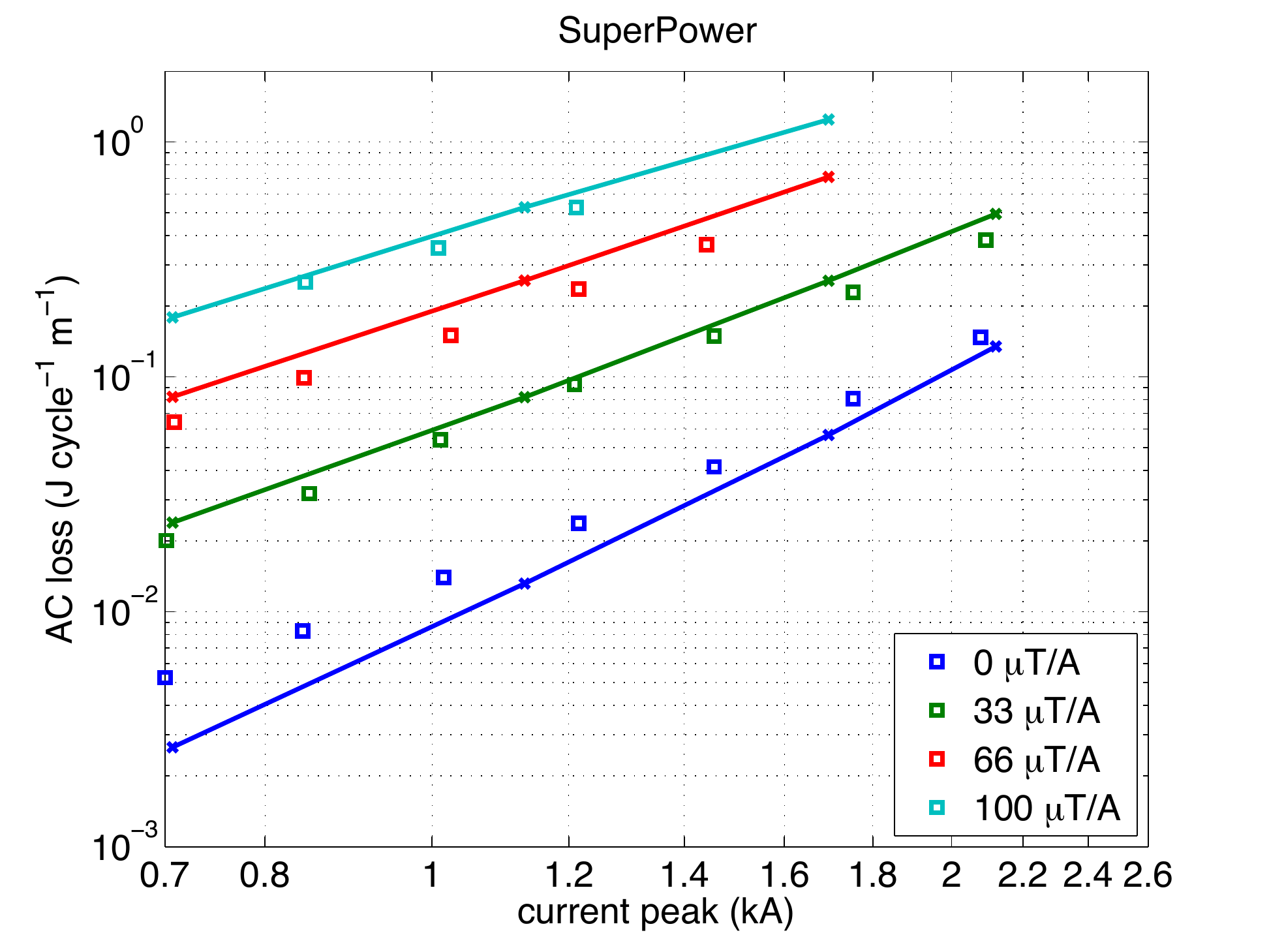}
\end{center}
\caption{\label{fig:TotalLosses_SuperPower}Measured and calculated AC losses of the SuperPower cable as a function of the transport current and for different background fields.}
\end{figure}

\section{Conclusion}
The focus of this paper was the estimation of self-field critical current and transport-magnetization AC losses of Roebel cables by means of numerical simulations.
Two cables with tapes from different manufacturers were considered in this study.
Two different models using the measured angular dependence of the critical current of the tapes composing the cables have been used to calculate the critical current of the cables. The obtained results agree with the experimentally measured values within \SI{9}{\percent}. 

In addition, this paper presented for the first time simulation and experimental results of AC losses in Roebel cables under the simultaneous action of transport current and magnetic field, with the field varying proportionally to the current. This is a condition typically met in superconducting windings.
The numerical calculations reveal peculiar patterns of the current and field distributions resulting from the simultaneous action of transport current and external field. The calculated losses agree with those measured with calorimetric technique within \SI{40}{\percent}.

\bibliographystyle{IEEEtran}


\end{document}